\documentclass[a4paper,11pt]{article}
\pdfoutput=1 

\usepackage{jcappub} 

\usepackage[T1]{fontenc} 
\usepackage{hyperref}
\usepackage{amssymb}
\usepackage{multirow}

\title{\boldmath Impact of non-zero strange quark mass $(m_{s}\neq0)$ in $f(R,T)$ gravity admitting observational results of strange stars}

\author[1]{Debadri Bhattacharjee}
\author[2]{and Pradip Kumar Chattopadhyay}

\affiliation[1,2]{IUCAA Centre for Astronomy Research and Development (ICARD), Department of Physics, Cooch Behar Panchanan Barma University, Vivekananda Street, District: Cooch Behar, \\ Pin: 736101, West Bengal, India}

\emailAdd{debadriwork@gmail.com}
\emailAdd{ pkc$_{-}$76@rediffmail.com}

\abstract{In this article we propose a new class of isotropic strange star using Buchdahl-I metric ansatz in the context of MIT bag model equation of state considering of non-zero strange quark mass $(m_{s})$ in the framework of modified $f(R,T)$ theory of gravity. The barotropic form of MIT bag model equation of state and a specific class of $f(R,T)$ model, {\it viz.}, $f(R,T)=R+2\alpha_{c}T$ where $\alpha_{c}$ is termed as the gravity-matter coupling constant, produces a tractable set of solutions of Einstein field equations. From the allowed numerical values of the coupling constant $(\alpha_{c})$, we have considered a range of $\alpha_{c}$ from -2.0 to 2.0. Maximum mass and radius in this model is found by numerically solving the TOV equations and we note that within the stability window imposed by energy per baryon, for an arbitrary choice of bag constant $B=70~MeV/fm^{3}$, $m_{s}$ and $\alpha_{c}$ act as a constraining factor. Interestingly, the increment of $m_{s}$ and $\alpha_{c}$ results in a softer equation of state which leads to the decrease in the maximum mass and radius while negative values of $\alpha_{c}$ leads to a stiffer equation of state thereby increasing the maximum mass and radius in the present model. For physical application, we consider EXO 1745-248 and study the effects of $m_{s}$ and $\alpha_{c}$ on its radius. Using the formalism, we have analysed the characteristic properties of EXO 1745-248. Apart from that, we have predicted the radii of a wide range of strange star candidates in the context of $f(R,T)$ gravity and the obtained results agree well with the observed results. We note that the proposed model satisfies all the necessary energy conditions and stability criteria to emerge as a viable stellar configuration.}

\begin{document}
\maketitle
\flushbottom

\section{Introduction}\label{sec1} Ever since Einstein unveiled his revolutionary theory of General Relativity (GR), gravity has been cast as the master weaver of spacetime. In 1915, GR was introduced and it was experimentally verified by Eddington in 1920. Despite its marvelous achievements in explaining the sectors where Newtonian theory failed, there are still many unresolved issues that require modifications of GR for the necessary explanations. Soon after the discovery, GR was first modified by Weyl \cite{Weyl} in 1919, through the introduction of higher order invariants in the Einstein-Hilbert action to unify electromagnetism and gravitation. Since then many researchers have imposed modifications on GR to explain various astrophysical and cosmological aspects. In the recent past, the theory of gravity has faced major fundamental challenge due to the observational results \cite{Riess,Perlmutter,Bernardis,Hanany,Peebles,Padmanabhan} on the accelerating phase of the present universe and possible existence of dark matter. An interesting approach to explain the observations is the consideration that GR breaks down at large scales and we need a more generalised action for the gravity field. In this purpose, many theoretical models have been introduced as extensions of GR, namely, $f(R)$, $f(T)$, $f(G)$, $f(R,T)$, $f(G,T)$ gravity etc. The simplest procedure amongst these extensions is $f(R)$ theory of gravity \cite{Nojiri,Sotiriou,Felice,Nojiri1,Carroll} where the standard Einstein-Hilbert action is replaced by an arbitrary function of Ricci scalar $R$. $f(R)$ theory has successfully improved our understanding of the galactic dynamics and accelerating universe \cite{Capozziello,Borowiec,Martins,Boehmer,Boehmer1,Cognola,Navarro,Song}. Harko et al. \cite{Harko} further generalised the $f(R)$ model by introducing the trace of energy-momentum tensor $T$ as a new component and presented the $f(R,T)$ gravity theory. In this pioneering work, the authors have mentioned that $T$ dependence points towards the quantum effects or exotic imperfect fluids \cite{Harko1}. In this study they have showed the non-nullification of covariant derivative of the energy-momentum tensor and the pervasive additional acceleration component emerging from the existence of matter-curvature coupling. Therefore, in essence, the test particles will follow a non-geodesic path in $f(R,T)$ gravity. Later, Chakraborty \cite{Chakraborty} showed that for a specific form $f(R,T)=R+hT$, the test particles do follow a geodesic path. An interesting aspect of $f(R,T)$ gravity is that it can provide an apt classical description of quantum gravity properties. In the cosmological scenario, there are vast motivations for using $f(R,T)$ models such as holographic dark energy, solar system consequences, anisotropic cosmology, non-equilibrium thermodynamics etc. However, studying these modified gravity theories using relativistic stellar models is vital in shaping a suitable theory of gravity because some explanations of these modifications are based on the idea that in the strong gravity regime, relativistic stars can differentiate between the gravitational laws and the major generalisations. Considering all these facts, we consider $f(R,T)$ gravity from the set of GR extensions in the present work. 
\\
\\
In literature, extensive research has been done on this domain. But in the last few years, there has been a particular surge in using $f(R,T)$ gravity in relativistic stellar modeling though the functional form $f(R,T)=R+2\alpha_{c}T$ \cite{Harko}, where $\alpha_{c}$ is known as the coupling constant, which we have adopted in this present study. In the context of neutron star and strange stars, Moraes et al. \cite{Moraes4}, first presented the exact solution of Tolman-Oppenheimer-Volkoff \cite{Tolman,Oppenheimer} (TOV) equation in $f(R,T)$ gravity and further analysed the hydrostatic equilibrium. Using the results obtained by Moraes et al. \cite{Moraes4}, Das et al. \cite{Das}, described compact stars in the framework of $f(R,T)$ gravity by employing Lie algebra and conformal Killing vectors. Considering $f(R,T)$ as linear functions of $R$ and $T$ along with embedding class-I approach, Errehymy et al. \cite{Errehymy} constructed anisotropic compact star models. Sarkar et al. \cite{Sarkar} have studied spherically symmetric anisotropic compact star solutions in $f(R,T)$ theory considering a suitable form of $g_{rr}$ metric potential. Sharif and Yousaf \cite{Sharif1} have studied the dynamical factors affecting the stability of an isotropic and spherically symmetric stellar configuration in $f(R,T)$ gravity. Yadav et al. \cite{Yadav} constructed singularity free solution in the context of non-exotic compact stars in $f(R,T)$ gravity. Rej and Bhar \cite{Rej} analysed isotropic compact stars in $f(R,T)$ gravity employing Durgapal IV metric potential. Using Buchdahl ansatz in the framework of $f(R,T)$ gravity, Kumar et al. \cite{Kumar1} studied an isotropic stellar model. Apart from that there have been many articles demonstrating different aspects of relativistic compact objects in $f(R,T)$ gravity \cite{Bhar,Lobato,Maurya,Hansraj,Prasad,Azmat,Hansraj1}. 
\\
\\
Compact astrophysical objects are classified into different sub classes on the basis of their initial mass content in the main sequence phase. One of such classes is Neutron Star (NS). NS have a layered structure with a high density core region $(\sim 2.7\times10^{14}~g/cm^{3})$. This high density regime is a fantastic laboratory to test the complexities in GR. In the theoretical perspective, it is imperative to study the astrophysical compact objects to obtain their mass and radii. Oppenheimer and Volkoff \cite{Oppenheimer} obtained a $0.7~M_{\odot}$ mass limit of NS considering the neutrons as ideal Fermi gas. Determination of maximum mass in compact stellar bodies depends on the solution of hydrostatic equilibrium which in turn shows that the maximum mass of any compact object is highly equation of state (EoS) dependent. Till date, the exact EoS that defines the interior matter distribution in NS is unknown. Various predictions have been made to find the maximum mass limit of compact objects. Rhoades and Ruffini \cite{Rhoades} in their study considered a perfect interior fluid distribution having density above the nuclear density range and within the causal limit they obtained the maximum mass of NS to be $3.2~M_{\odot}$. Nauenberg and Chapline \cite{Nauenberg} further extended the maximum mass to $3.6~M_{\odot}$. In the context of a non-rotating NS having EoS above the nuclear density range, Sabbadini and Hartle \cite{Sabbadini} obtained a maximum mass of $5~M_{\odot}$ without considering the causality conditions. Different assumptions of compact object EoS reveal a wide range of maximum mass $1.46$-$2.48~M_{\odot}$ \cite{Haensel}. Over the years, Researchers have shown a great interest in studying the internal composition of the compact stars in different perspectives. Quark star hypothesis, where stars are hypothesised to be composed of strange quark matter (SQM), is one such theories that has emerged as a prime candidate to describe the NS interior structure based on the recent observational aspects. At first, it was Itoh \cite{Itoh} who suggested that quark stars can be found to be in a state of hydrostatic equilibrium. Considering zero external pressure, Madsen \cite{Madsen} showed that quark stars containing only $u$ and $d$ quarks are unstable. On the other hand, inclusion of $s$ quark reduces the energy per baryon of the system effectively on comparison to a two flavour system. Therefore, the inclusion of $s$ quarks are necessary. Based on this theory, a class of compact objects are grouped together and they are called Strange Stars (SS) or Strange Quark Stars (SQS) \cite{Baym, Alcock1,Madsen,Glendenning}. Witten \cite{Witten} introduced that idea that Strange Quark Matter (SQM) may be termed as the true ground state of quantum chromodynamics (QCD) as the associated energy per baryon for SQM is less than that of the most stable nuclei $^{56}Fe$. This proposition supports further stability of SQM. The asymptotic freedom in QCD dictates that in the high energy regime the nucleons dissolve into their constituent quarks which creates a weakly interacting quark-gluon plasma (QGP) phase. Alford \cite{Alford} proposed that the high density and low temperature regime of NS core is sufficient for the formation of quark matter.  While our understanding of the hadron to quark phase transition  remains under development, a variety of phenomenological models offer insights into this process. One such model is the MIT bag model \cite{Chodos}. This model describes the behaviour of bulk quark matter and their deconfined state in a spatial region called \lq bag', which is effectively the difference of perturbative and non-perturbative vacuum, through the EoS  
\begin{equation}
	p=\frac{1}{3}(\rho-4B), \label{eq1}
\end{equation}
where $p$ is the pressure, $\rho$ is the energy density and $B$ is the bag constant. In the MIT bag model, quarks are considered as degenerate Fermi gas of $u$, $d$, massive $s$ quarks and electrons. In the framework of MIT bag model, Farhi and Jaffe \cite{Farhi} have explored the effect of mass of strange quark $(m_{s})$ on the gross properties of SQM. In their study, they obtained a stability window for stable SQM in the $m_{s}-B$ space. Using the MIT bag model in GR, many researchers have extensively studied the SQS properties \cite{Brilenkov,Paulucci,Arbanil,Lugones,Chowdhury,KBG,KBG1,Hakim,Abbas} and so on. Interestingly, in recent times, the inclusion of modified theories of gravity in the SQS modeling has revealed some intriguing features about compact stellar modeling. In the context of $f(R,T)$ gravity, Deb et al. \cite{Deb} studied the SS considering Mak-Harko \cite{Mak} density profile and MIT bag model EoS. Biswas et al. \cite{Biswas} have studied singularity free solutions of SS in $f(R,T)$ gravity using Krori-Barua \cite{KB} metric potential. Carvalho et al. \cite{Carvalho1} have numerically solved the TOV equation \cite{Tolman,Oppenheimer} using MIT bag model EoS to obtain a maximum mass for SS. Several authors have studied the SS modeling in the framework of $f(R,T)$ gravity \cite{Deb1,Sharif2,Noureen1,Maurya1}. 
\\ 
\\
The choice of metric ansatz plays a vital role in obtaining a tractable set of exact solutions of Einstein Field Equations (EFE). Considering static equilibrium of perfect fluid distribution Delgaty and Lake \cite{Delgaty} created a catalogue of spherically symmetric solutions, be it exact or closed, of EFE. These solutions are very important for the determination of viable compact object properties. The Buchdahl-I metric ansatz \cite{Buchdahl} is one such metric potentials which has proven to be very useful in the study of compact astrophysical objects. For analytical modeling of relativistic fluid sphere, Durgapal and Banerji \cite{Durgapal} rederived the Buchdahl-I ansatz. In another study, Maurya et al. \cite{Maurya2} used the Buchdahl-I metric ansatz to explore the hydrostatic equilibrium of compact stars in $f(R,T)$ gravity. Recently, Bhattacharjee and Chattopadhyay \cite{Bhattacharjee1} have used the metric ansatz to study the maximum mass of compact stars using modified Chaplygin gas EoS. 
\\
\\ 
The intriguing aspect of the inclusion of SQM in $f(R,T)$ gravity in the context of SQS modeling is the main motivation for the present paper. The remainder of the paper is organised in the following manner: In Section~\ref{sec2}, we have described the basic thermodynamics at T$\rightarrow$ 0 K arising due to the inclusion of SQM and the subsequent modification of MIT bad model EoS. Here, we have also defined the different stability window for different choices of bag constant $(B)$ and mass of strange quarks $(m_{s})$. Section~\ref{sec3} address the mathematical formalism of $f(R,T)$ theory. In Section~\ref{sec4}, considering spherically symmetric line element, we construct and solve the EFE in the framework of $f(R,T)$ gravity and obtain the exact solutions for the energy density $(\rho)$, pressure $(p)$ and $g_{tt}$ component. Next, Section~\ref{sec5} deals with the boundary condition where we have matched the interior and exterior solutions. Section~\ref{sec6} gives the necessary bounds on the coupling parameter $(\alpha_{c})$. The effect of coupling parameter $(\alpha_{c})$ and strange quark mass $(m_{s})$ on the maximum mass and radius is depicted in Section~\ref{sec7}. Physical application of the model is done in Section~\ref{sec8} along with the causality and energy conditions. The stability of the model is assessed on the basis of generalised TOV equations, adiabatic index, lagrangian perturbation in Section~\ref{sec9}. Tidal love number and tidal deformability are evaluated in this section also. Finally, in Section~\ref{sec10}, we discuss the major findings of the paper.        
\section{Thermodynamics of strange quark matter at T$\rightarrow$ 0 K}\label{sec2} 
We consider the interior matter distribution of a highly dense strange quark star composed of by quark matter governed by the MIT bag model EoS \cite{Chodos}. Quarks, being Fermi particles, the interior may be considered to be composed of degenerate Fermi gas of quarks where quarks are in the de-confined phase. If A is the baryon number, then the quark matter signifies a Fermi gas of 3A quarks constituting a single colour-singlet. Following MIT bag model, the dynamics of quark confinement is approximated as \cite{Chodos,Kettner}:
\begin{equation}
	p+B=\sum_{i=u,d,s,e^{-}}p_{i}, \label{eq2}
\end{equation}   
and 
\begin{equation}
	\rho-B=\sum_{i=u,d,s,e^{-}}\rho_{i}, \label{eq3}
\end{equation}  
where, $p_{i}$ and $\rho_{i}$ are the pressure and energy density of the ith particles and $B$ is the bag constant. In this present scenario, we consider the strange matter to be made up of three types of quarks, namely $u$, $d$, $s$ quarks and electrons $(e^{-})$ only. Following Kettner et al. \cite{Kettner}, we infer that the possible existence of charm quark stars is eliminated as they are unstable under radial oscillations. Moreover, as the population density of muons becomes significant above a threshold charm quark density, hence we can neglect the muons as well \cite{Kettner}. In case of SQS, coulomb interaction dominates over the gravity effects. As a consequence, at the possibly lowest energy state, SQM is considerably charge neutral \cite{Alcock}. The charge neutrality condition is written as: 
\begin{equation}
	\sum_{i=u,d,s,e^{-}}n_{i}q_{i}=0, \label{eq4}
\end{equation} 
where, the subscript \lq i\rq~stands for the i-th particle. 
As the chemical potential of the quarks are greater than the temperature of the star, we can approximate the expressions for the quark pressure $(p_{i})$, energy density $(\rho_{i})$ and number density $(n_{i})$ in the case $T\rightarrow0$ as \cite{Kettner,Peng}: 
\begin{equation}
	\hspace{-1cm}
	p_{i}=\frac{g_{i}\mu_{i}^{4}}{24\pi^{2}}\sqrt{1-\Big({\frac{m_{i}}{\mu_{i}}}\Big)^{2}}\Big\{{1-\frac{5}{2}\Big(\frac{m_{i}}{\mu_{i}}\Big)^{2}}\Big\}+\frac{3}{2}\Big(\frac{m_{i}}{\mu_{i}}\Big)^{4}\ln{\frac{1+\sqrt{1-\Big(\frac{m_{i}}{\mu_{i}}\Big)^{2}}}{\Big(\frac{m_{i}}{\mu_{i}}\Big)}}, \label{eq5}
\end{equation} 
\begin{equation}
	\hspace{-1cm}
	\rho_{i}=\frac{g_{i}\mu_{i}^{4}}{24\pi^{2}}\sqrt{1-\Big({\frac{m_{i}}{\mu_{i}}}\Big)^{2}}\Big\{{1-\frac{1}{2}\Big(\frac{m_{i}}{\mu_{i}}\Big)^{2}}\Big\}-\frac{1}{2}\Big(\frac{m_{i}}{\mu_{i}}\Big)^{4}\ln{\frac{1+\sqrt{1-\Big(\frac{m_{i}}{\mu_{i}}\Big)^{2}}}{\Big(\frac{m_{i}}{\mu_{i}}\Big)}}, \label{eq6}
\end{equation}
\begin{equation}
	\hspace{-1cm}
	n_{i}=\frac{g_{i}\mu_{i}^{3}}{6\pi^{2}}\Big[1-\Big(\frac{m_{i}}{\mu_{i}}\Big)^{2}\Big]^{\frac{3}{2}}. \label{eq7}
\end{equation}
For all flavour of quarks, the degeneracy factor $g_{i}=6$ and for electrons $g_{i}=2$. The charge neutrality condition as a written in eq.~\eqref{eq4} is modified in the following form: 
\begin{equation}
	2\Big(1-\frac{\mu_{e^{-}}}{\mu}\Big)^{3}-\Big(\frac{\mu_{e^{-}}}{\mu}\Big)^{3}-\Big\{1-\Big(\frac{m_{i}}{\mu_{i}}\Big)^{2}\Big\}^{\frac{3}{2}}-1=0, \label{eq8}
\end{equation} 
where, $\mu=\mu_{d}=\mu_{s}$ is the chemical potential of the $d$ and $s$ quarks respectively. Generally, in the context of SQS being composed $u$, $d$ and $s$ quarks, $u$ and $d$ quarks are much lighter than the $s$ quarks and the mass of the $s$ quarks are assumed to be zero $(m_{s}\rightarrow0)$. Using $m_{s}\rightarrow0$ in the modified charge neutrality condition, we find that $\mu_{e^{-}}\rightarrow0$. This in turn shows that the electrons are not necessarily important in maintaining the charge neutrality of a massless quark system. Interestingly, under this condition we can construct the EoS $p_{i}=\frac{1}{3}\rho_{i}$ from eqs.~\eqref{eq4} and \eqref{eq5}. Using eqs.~\eqref{eq2} and \eqref{eq3}, for a system of massless strange quarks, we can establish the MIT bag model EoS as \cite{Kapusta}: 
\begin{equation}
	p=\frac{1}{3}(\rho-4B). \label{eq9}
\end{equation}
The inclusion of finite strange quark mass $(m_{s}\neq0)$ in our analysis along with eqs.~\eqref{eq2}, \eqref{eq3}, \eqref{eq5}, \eqref{eq6} lead to the most general form of the EoS describing SQM which is expressed as: 
\begin{equation}
	\rho=3p+4B+\rho_{s}-3p_{s}, \label{eq10} 
\end{equation}
\begin{equation}
	\Rightarrow p=\frac{1}{3}(\rho-4\tilde{B}), \label{eq11}
\end{equation}
where, $\tilde{B}=B+\frac{1}{4}(\rho_{s}-3p_{s})$, $\rho_{s}$ is the density of the strange quark and $p_{s}$ is the pressure exerted due to the presence of strange quark. The energy per baryon $(\mathcal{E_{B}})$ for the most stable nuclei, i.e., $^{56}Fe$ is $930.4~MeV$. Therefore, energy per baryon for a quark matter made of two flavour quarks is greater than this value, otherwise, $^{56}Fe$ would have been made of two flavour quarks namely, $u$ and $d$ quarks, which in reality has not been observed. Therefore, three flavour quark system is considered to be stable when the energy per baryon of the configuration is less than $930.4~MeV$. The metastability condition constrains the energy per baryon within the limit $940.4~MeV<\mathcal{E_{B}}<939~MeV$ \cite{Backes} which is in the typical nucleonic mass range. Above this range, i.e., $\mathcal{E_{B}}>939~MeV$, the SQM becomes unstable. The baryon number density for the three-flavour SQM is expressed as: 
\begin{equation}
	n_{B}=\frac{1}{3}\sum_{i=u,d,s}n_{i}. \label{eq12}
\end{equation} 
\begin{figure}[ht!]
	\centering
	\includegraphics[width=0.5\textwidth]{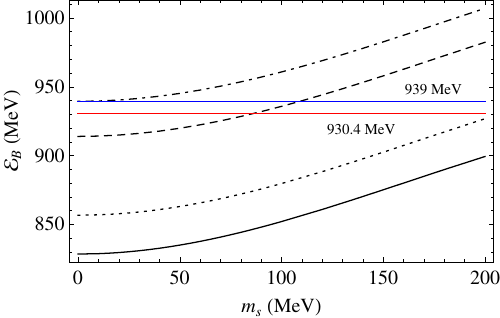}
	\caption{Variation of energy per baryon $(\mathcal{E_{B}})$ with strange quark mass $(m_{s})$ for different bag value $(B)$. The solid, dotted, dashed and dot-dashed lines represent $B=57.55,~65.80,~85.21$ and $95.11~MeV/fm^{3}$ respectively.}
	\label{fig1} 
\end{figure}
In figure~(\ref{fig1}), we have plotted the variation of energy per baryon $(\mathcal{E_{B}})$ with strange quark mass $(m_{s})$ for different bag value $(B)$. From the previous arguments, it is established that the energy per baryon $(\mathcal{E_{B}})$ depends on the bag value $(B)$ as well as the mass of the strange quarks $(m_{s})$. Accordingly, depending on particular combinations of $B$ and $m_{s}$, the three flavour quark matter may be stable, metastable or unstable. From figure~(\ref{fig1}), we have obtained a constrained range of $m_{s}$ for which the strange matter should be metastable or unstable. We have tabulated the range of $m_{s}$  in table~(\ref{tab1}) for different values of $B$ considered in figure~(\ref{fig1}). 
\begin{table}[h!]
	\centering
	\vspace{-0.4cm}
	\begin{tabular}{ccccc}
		\hline
		\multirow{2}{*}{ Stability }& $B=57.55$ & $B=65.80$ & $B=85.21$ & $B=91.55$ \\ 
		& $(MeV/fm^{3})$ & $(MeV/fm^{3})$ & $(MeV/fm^{3})$ & $(MeV/fm^{3})$ \\ \hline
		Stable & \multirow{2}{*}{$m_{s}<290~MeV$} & \multirow{2}{*}{$m_{s}<207~MeV$}&  \multirow{2}{*}{$m_{s}<86~MeV$} & \multirow{2}{*}{....}\\ 
		$(\mathcal{E_{B}}<930.4~MeV)$ &&&& \\ \hline
		MetaStable & \multirow{2}{*}{....} & \multirow{2}{*}{....} & \multirow{2}{*}{$86<m_{s}<108.1~MeV$} & \multirow{2}{*}{....}\\
		$(930.4<\mathcal{E_{B}}<939~MeV)$ &&&& \\\hline
		Unstable & \multirow{2}{*}{....} & \multirow{2}{*}{....} & \multirow{2}{*}{$m_{s}>108.1~MeV$} & \multirow{2}{*}{$m_{s}>0~MeV$}\\
		$(\mathcal{E_{B}}>939~MeV)$ &&&& \\ \hline
	\end{tabular}
	\caption{Stability window of 3 flavour SQM with constrained $m_{s}$ and different bag value $(B)$.\label{tab1}} 
\end{table}  
To ensure a stable strange star modeling, we have considered the value of $B=70~MeV/fm^{3}$ from table~(\ref{tab1}). 
\section{Mathematical formalism of $f(R,T)$ gravity}\label{sec3}
The total Einstein-Hilbert action for the $f(R,T)$ gravity theory can be written as \cite{Harko}:
\begin{equation}
	S_{EH}=\frac{1}{4\pi}\int d^{4}xf(R,T)\sqrt{-g}+\int d^{4}xL_{m}\sqrt{-g}, \label{eq13}
\end{equation}
where $R$ is the Ricci Scalar and $T$ is the trace of the energy-momentum tensor $(T_{ij})$, $L_{m}$ is the matter Lagrangian density and $g=det(g_{ij})$ is the fundamental metric determinant, here we have assumed the relativistic units $G=c=1$. The Lagrangian matter density function is associated with the energy-momentum tensor as:
\begin{equation}
	T_{ij}=-\frac{2}{\sqrt{-g}}\frac{\delta (\sqrt{-g}L_{m})}{\delta g^{ij}}. \label{eq14}
\end{equation} 
On the other hand, if $L_{m}$ is solely dependent on $g_{ij}$, then eq.~\eqref{eq14} can be written as \cite{Landau}:
\begin{equation}
	T_{ij}=g_{ij}L_{m}-2\frac{\partial L_{m}}{\partial g^{ij}}. \label{eq15}
\end{equation}
The variation of the action in eq.~\eqref{eq11} with respect to the fundamental metric tensor $g_{ij}$ yields the modified form of field equation in $f(R,T)$ in the following form:
\begin{eqnarray}
	f_{R}(R,T)R_{ij}-\frac{1}{2}f(R,T)g_{ij}+(g_{ij}\Box-\nabla_{i}\nabla_{j})f_{R}(R,T)\nonumber\\\hspace{2cm}=8\pi T_{ij}-f_{T}(R,T)T_{ij}-f_{T}(R,T)\Theta_{ij}, \label{eq16}
\end{eqnarray}
where, $f_{R}(R,T)=\frac{\partial f(R,T)}{\partial R}$ and $f_{T}(R,T)=\frac{\partial f(R,T)}{\partial T}$, $R_{ij}$ is the Ricci tensor, $\Box$ defines the D' Alembert operator, $\Box\equiv~\frac{\partial_{i}(\sqrt{-g}g^{ij}\partial_{j})}{\sqrt{-g}}$. $\nabla_{j}$ represents the covariant derivative related to Levi-Civita connection of $g_{ij}$, $\Theta=g^{\alpha\beta}\frac{\delta T_{\alpha\beta}}{\delta g_{ij}}$.
The covariant divergence of eq.~\eqref{eq15} yields \cite{Barrientos}:
\begin{equation}
	\hspace{-2cm}	\nabla^{i}T_{ij}=\frac{f_{T}(R,T)}{8\pi-f_{T}(R,T)}\Big[(T_{ij}+\Theta_{ij})\nabla^{i}~ln~f_{T}(R,T)+\nabla^{i}\Theta_{ij}-\frac{1}{2}g_{ij}\nabla^{i}T\Big]. \label{eq17}
\end{equation} 
Eq.~\eqref{eq16} shows the non-conservation of energy-momentum tensor in $f(R,T)$ gravity. \\
In this formalism, we consider the energy-momentum tensor of an isotropic perfect fluid distribution as:
\begin{equation}
	T_{ij}=(p+\rho)u_{i}u_{j}+pg_{ij}, \label{eq18}
\end{equation} 
where, $p$ is the isotropic pressure, $\rho$ is the energy density and $u^{i}u_{i}=-1$ is the four velocity vector respectively. Following Ref. \cite{Pretel}, corresponding to the perfect fluid distribution, we consider the matter Lagrangian as $L_{m}=p$. Previously, Faraoni \cite{Faraoni} demonstrated that, for minimal fluid-gravity coupling, $L_{m}=p$ and $L_{m}=-\rho$ are both analogous. Moreover, $\rho$ and $p$ being freely interchangeable thermodynamic quantities, Harko \cite{Harko1} showed that expressing the Lagrangian through $\rho$ and $p$ are completely equivalent. Accordingly, $\Theta_{ij}=-2T_{ij}+pg_{ij}$, $\Theta=-2T+4p$. \\
In this present study, following Harko et al. \cite{Harko}, we focus on a specific class of $f(R,T)$ modified gravity models, namely, $f(R,T)=R+2\alpha_{c}T$, where $\alpha_{c}$ is the coupling parameter. To ensure realistic modeling, $\alpha_{c}$ must have some restrictions considering the astrophysical as well as cosmological scales. In this regard, Lobato et al. \cite{Lobato} showed that due to the presence of neutron star crust, $|\alpha_{c}|$ should be small. From dark energy density parameter, it was established in Ref. \cite{Bhattacharjee} that $\alpha_{c}\gtrsim-1.9\times10^{-8}$. Stellar equilibrium of white dwarfs show $\alpha_{c}>-3.0\times10^{-4}$ \cite{Carvalho}. In the context of background evolution, Velten and Caram\^es \cite{Velten} restricted $\alpha_{c}$ as $-0.1<\alpha_{c}<1.5$. In literature, this particular functional form of $f(R,T)$ gravity \cite{Shabani,Shabani1,Singh,Moraes,Moraes1,Moraes2,Moraes3,Sharif,Reddy,Kumar,Shamir} is used to develope many successful models. Moreover, several authors have constructed astrophysical models in the framework of $f(R,T)$ gravity using positive values of coupling parameter $(\alpha_{c})$ \cite{Das,Sarkar,Sharif1,Yadav,Bhar,Biswas,Noureen,Zubair} while the negative range has been considered in Refs. \cite{Deb,Carvalho1}. Substitution of the chosen form of $f(R,T)$ in eq.~\eqref{eq16} leads to the modified form of Einstein field equation as:
\begin{equation}
	G_{ij}=8\pi T_{ij}+\alpha_{c}Tg_{ij}+2\alpha_{c}(T_{ij}-pg_{ij}), \label{eq19}
\end{equation} 
where, $G_{ij}$ is the Einstein tensor and we retain the field equations of GR in the case of $\alpha_{c}=0$. Consequently, the covariant divergence of the energy-momentum tensor given in eq.~\eqref{eq16} reduces to th following form:
\begin{equation}
	\nabla^{i}T_{ij}=\frac{2\alpha_{c}}{8\pi+2\alpha_{c}}\Big[\nabla^{i}(pg_{ij})-\frac{1}{2}g_{ij}\nabla^{i}T\Big]. \label{eq20}
\end{equation} 
\section{Einstein Field Equations and their solutions in $f(R,T)$ gravity}\label{sec4} In the curvature co-ordinate system $(t,r,\theta,\phi)$ 
\begin{equation}
	ds^2=-e^{2\nu(r)}dt^2+e^{2\lambda(r)}dr^2+r^2(d\theta^2+sin^2\theta d\phi^2) \label{eq21}
\end{equation}
corresponds to a static, spherically symmetric line element where $\nu$ and $\lambda$ are radial functions respectively. Substituting eq.~\eqref{eq18} in eq.~\eqref{eq19}, we deduce a tractable set of the modified form of Einstein field equations as:
\begin{eqnarray}
	\frac{2\lambda'e^{-2\lambda}}{r}+\frac{1-e^{-2\lambda}}{r^2}=8\pi\rho-\alpha_{c}(p-3\rho), \label{eq22} \\
	\frac{2\nu'e^{-2\lambda}}{r}-\frac{1-e^{-2\lambda}}{r^2}=8\pi p+\alpha_{c}(3p-\rho), \label{eq23} \\
	e^{-2\lambda}(\nu''+\nu'^{2}-\lambda'\nu'+\frac{\nu'}{r}-\frac{\lambda'}{r})=8\pi p+\alpha_{c}(3p-\rho), \label{eq24}
\end{eqnarray}
where, $(')$ denotes the derivative with respect to $r$. Using eqs.~\eqref{eq22}, \eqref{eq23} and \eqref{eq24} we obtain, 
\begin{equation}
	p'+(p+\rho)\nu'=\frac{\alpha_{c}}{8\pi+2\alpha_{c}}(\rho'-p'). \label{eq25}
\end{equation}
Interestingly, for $\alpha_{c}=0$, from eq.~\eqref{eq25}, we retain the conservation equation of Einstein gravity. 

To obtain the exact set of solutions in this formalism, we introduce the Buchdahl-I ansatz \cite{Buchdahl} encompassing almost every solution of the static EFE coupled with a perfect fluid distribution and it is expressed as: 
\begin{equation}
	e^{2\lambda}=\frac{2(1+\eta r^{2})}{2-\eta r^{2}}, \label{eq26}
\end{equation} 
where, $\eta$ is a constant with dimension $Km^{-2}$. Using eq.~\eqref{eq26} in eqs.~\eqref{eq22}, \eqref{eq23} and \eqref{eq24}, we obtain the expression for the energy density as:
\begin{equation}
	\rho=\frac{27\eta+9\eta^{2}r^{2}-8\tilde{B}\alpha_{c}-16\tilde{B}\eta\alpha_{c}r^{2}-8\tilde{B}\alpha_{c}\eta^{2}r^{4}}{16(3\pi+\alpha_{c})(1+\eta r^{2})^{2}}. \label{eq27}
\end{equation}
From eq.~\eqref{eq11}, pressure $(p)$ is expressed as:
\begin{equation}
	p=\frac{1}{3}\Big(\frac{27\eta+9\eta^{2}r^{2}-8\tilde{B}\alpha_{c}-16\tilde{B}\eta\alpha_{c}r^{2}-8\tilde{B}\alpha_{c}\eta^{2}r^{4}}{16(3\pi+\alpha_{c})(1+\eta r^{2})^{2}}-4\tilde{B}\Big). \label{eq28}
\end{equation}
Now, using eqs.~\eqref{eq26}, \eqref{eq27} and \eqref{eq28} in eq.~\eqref{eq23}, we obtain:
\begin{equation}
	\nu=\frac{log(2-\eta r^{2})\Big[24\tilde{B}(8\pi^{2}+6\pi\alpha_{c}+\alpha_{c}^{2})-\eta(3\alpha_{c}+14\pi)\Big]+2\eta\Big[4\tilde{B}r^{2}(8\pi^{2}+6\pi\alpha_{c}+\alpha_{c}^{2})+\pi log(1+\eta r^{2})\Big]}{4\eta(3\pi+\alpha_{c})}. \label{eq29}
\end{equation}
Here, without any loss of generality, we have taken the constant of integration to be zero. 
The total active gravitational mass contained within the sphere of radius $R$ is written as:
\begin{equation}
	m(r)=\int_{0}^{R}4\pi r^{2}\rho dr. \label{eq30}
\end{equation}
\section{Boundary Conditions}\label{sec5} 
It has been well established that the matching conditions of GR undergo modifications in presence of modified theories of gravity and extra scalar dimensions \cite{Rosa,Rosa1}. Following the formalism described in Ref.~\cite{Rosa1}, we substitute the specific form of $f(R,T)=R(1+\gamma T)-2\Lambda+\sigma R^{2}+h(T)$ in eq.~\eqref{eq16} and we obtain:
\begin{eqnarray}
	(1+\gamma T+2\sigma R)R_{ij}-\frac{1}{2}g_{ij}[R(1+\gamma T)-2\Lambda+\sigma R^{2}+h(T)]-\nonumber\\(2\sigma R+\gamma T)(\nabla_{i}\nabla_{j}-g_{ij}\Box)=\Bigg(8\pi+\frac{\partial h(T)}{\partial T}+\gamma R\Bigg)T_{ij}+\Bigg(\frac{\partial h(T)}{\partial T}+\gamma R\Bigg)pg_{ij}, \label{eq30a}
\end{eqnarray} 
where, $\gamma,~\Lambda,~\sigma$ are arbitrary constants. Now, to describe the generalised structure of the matching condition, we consider a hyper-surface $\Big(\sum\Big)$. It is imperative that we must preserve the continuity of $h(T)$ across $\sum$ to derive the matching conditions between the interior and the exterior space-times. Accordingly, one can verify that the energy-momentum tensor in this case must be written in the form:
\begin{equation}
	T_{ij}=T^{+}\Theta(\ell)+T^{-}\Theta(-\ell)+\delta(\ell)(S_{ij}+2S_{(_{i}n_{j})}+Sn_{i}n_{j})+s_{ij}(\ell), \label{eq30b}
\end{equation}
where, $(+)$ and $(-)$ represent the exterior and interior values respectively, $S_{ij}$ is the energy-momentum tensor at $\sum$, $S_{i}$ is the external contribution of flux momentum having a normal component that measures the normal energy energy flux across the hyper-surface ($\sum$), $S$ is the external normal pressure on $\sum$ and $s_{ij}$ is the energy-momentum tensor at the double-layer. Substituting eq.~\eqref{eq30b} in eq.~\eqref{eq30a} we obtain the form of junction condition as \cite{Rosa1}:
\begin{eqnarray}
	\Bigg(8\pi+\gamma R^{\sum}+\frac{\partial h(T^{\sum})}{\partial T^{\sum}}\Bigg)S_{ij}&=&-\Bigg(1+\gamma T^{\sum}+2\sigma R^{\sum}\Bigg)\epsilon[K_{ij}]+\nonumber\\&&\epsilon \hat{e_{ij}}n^{c}\Bigg(2\sigma[\nabla_{c}R]+\gamma[\nabla_{c}T]\Bigg)-\epsilon K^{\sum}_{ij}\Bigg(2\sigma[R]+\gamma[T]\Bigg), \label{eq30c}
\end{eqnarray}
where, $\hat{e_{ij}}=g_{ij}\mathfrak{e^{i}_{a}}\mathfrak{e^{j}_{b}}$ is the metric induced at the hyper-surface $\Big(\sum\Big)$ by the fundamental metric $g_{ij}$, $K_{ij}=\nabla_{i}n_{j}$ is the extrinsic curvature tenor at $\sum$ and $R^{\sum},~T^{\sum},~K^{\sum}_{ij}$ define the average values of these quantities at the hyper-surface $\Big(\sum\Big)$. Interestingly, in the present scenario, our choice of $f(R,T)=R+2\alpha_{c}T$ and a smooth matching condition $(S_{ij}=0)$ lead to the Darmois junction condition \cite{Darmois} of GR i.e., $[K_{ij}]=0$. Hence, despite using a specific model of $f(R,T)$ gravity, we can use the matching conditions of GR. To ensure a regular matching of the metric coefficients and to evaluate the specific constants appearing in the formalism, we match the interior space-time to the exterior vacuum Schwarzschild line element \cite{Schwarzschild} which is expressed as: 
\begin{equation}
	ds^2=-\Big(1-\frac{2M}{r}\Big)dt^2+\frac{1}{(1-\frac{2M}{r})}dr^2+r^2(d\theta^2+sin^2\theta d\phi^2). \label{eq31}
\end{equation}
The continuity of the metric potentials at the surface of the star $(r=R)$ yields, 
\begin{equation}
	e^{-2\lambda}=1-\frac{2M}{R}, \label{eq32} 
\end{equation}
and, 
\begin{equation}
	e^{2\nu}=1-\frac{2M}{R}. \label{eq33}
\end{equation}
The stellar boundary is defined as the surface where the pressure becomes zero, i.e, 
\begin{equation}
	p(r=R)=0. \label{eq34}
\end{equation}
Using eq.~\eqref{eq34}, we obtain the expression for stellar radius for the model as: 
\begin{equation}
	R=\frac{1}{4}\sqrt{\frac{3\eta^{2}-16\tilde{B}\eta(8\pi+3\alpha_{c})-\sqrt{3\eta^{3}(3\eta+64\tilde{B}(8\pi+3\alpha_{c}))}}{\tilde{B}\eta^{2}(8\pi+3\alpha_{c})}}. \label{eq35}
\end{equation}
Eq.~\eqref{eq35} shows the dependence of stellar radius on the model parameters. Now, using the eqs.~\eqref{eq32}, \eqref{eq33} and \eqref{eq34}, we find the expression for $\eta$ as:
\begin{equation}
	\eta= \frac{4M}{R^{2}(3R-4M)}. \label{eq36}
\end{equation}
\section{Bounds on the coupling parameter $\alpha_{c}$}\label{sec6} From eqs.~\eqref{eq27} and \eqref{eq28}, we can express the central energy density $(\rho_{0})$ and central pressure $(p_{0})$ as: 
\begin{equation}
	\rho_{0}= \frac{27\eta-8\tilde{B}\alpha_{c}}{48\pi+16\alpha_{c}}, \label{eq37}
\end{equation}
and, 
\begin{equation}
	p_{0}=\frac{9\eta-8\tilde{B}(8\pi+3\alpha_{c})}{16(3\pi+\alpha_{c})}. \label{eq38}
\end{equation}
Now, the positivity of the central pressure $(p_{0}>0)$ imposes a physical bound on the coupling parameter $(\alpha_{c})$ which is obtained as:
\begin{equation}
	-3\pi<\alpha_{c}<\frac{9\eta-64\pi\tilde{B}}{24\tilde{B}}. \label{eq39}
\end{equation}
It is evident from eqs.~\eqref{eq36} and \eqref{eq39} that the range of coupling parameter $\alpha_{c}$ is highly mass-radius specific. In this formalism, we have chosen the range of values of coupling parameter $(\alpha_{c})$ from -2.0 to 2.0 within the allowed range of $\alpha_{c}$ given in eq.~\eqref{eq39} and found that the central and surface energy density as well as central pressure remains positive for the range, hence the choice is apt. Similar range of coupling constant is also considered in Refs. \cite{Deb,Carvalho1}.
\section{Mass-Radius relationship from TOV equation}\label{sec7}
\begin{figure}[h]
	\centering
	\includegraphics[width=0.5\textwidth]{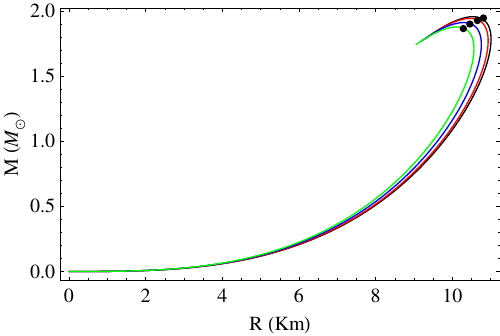}
	\caption{Mass-Radius plot for different $m_{s}$ keeping $B=70~MeV/fm^{3}$ and $\alpha_{c}=-1$. The black, red, blue and green lines correspond to $m_{s}=0,~50,~100,~150~MeV$ respectively.}
	\label{fig2}
\end{figure}

\begin{figure}[h]
	\centering
	\includegraphics[width=0.5\textwidth]{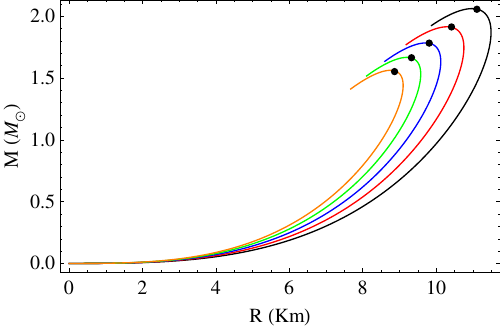}
	\caption{Mass-Radius plot for different coupling constant $\alpha_{c}$ keeping $B=70~MeV/fm^{3}$ and $m_{s}=100~MeV$. The black, red, blue, green and orange lines correspond to $\alpha_{c}=-2.0,~-1.0,~0,~1.0,~2.0$ respectively.}
	\label{fig3}
\end{figure}

In this section, we have solved the TOV equations \cite{Tolman,Oppenheimer} numerically to determine the maximum mass and the corresponding radius in this model. To ensure a stable stellar model, we have chosen $B=70~MeV/fm^{3}$. Figure~(\ref{fig2}) shows the mass-radius relationship when strange quark mass $(m_{s})$ is varied keeping $\alpha_{c}$ fixed at $-1.0$ and figure~(\ref{fig3}) shows the mass-radius relationship when the coupling parameter $(\alpha_{c})$ is varies keeping $m_{s}$ fixed at $100~MeV$. We note that, 
\begin{itemize}
	\item $m_{s}=0$ and $\alpha_{c}=0$ retains the MIT bag model in the context of GR. 
	\item the maximum mass decreases when we increase $m_{s}$ keeping $\alpha_{c}$ fixed and vice versa.
	\item the negative values of $\alpha_{c}$ allows larger values of maximum mass in the model.
\end{itemize}
The mass-radius values for figures~(\ref{fig2}) and (\ref{fig3}) are tabulated in tables~(\ref{tab2}) and (\ref{tab3}) respectively. 
\begin{table}[ht!]
	\centering
	\begin{tabular}{cccc}
		\hline
		Coupling parameter & Strange quark mass & Maximum mass & Radius\\
		$(\alpha_{c})$ & $(m_{s})~(MeV)$ & $(M_{\odot})$ & $(Km)$ \\ \hline
		\multirow{4}{*}{-1.0} & 0 & 1.96 & 10.54\\
		& 50 & 1.94 & 10.47 \\
		& 100 & 1.91 & 10.30 \\
		& 150 & 1.88 & 10.11 \\ \hline   
	\end{tabular}
	\caption{Maximum mass-radius for $B=70~MeV/fm^{3}$ and $\alpha_{c}=-1.0$.\label{tab2}}
\end{table} 
\begin{table}[ht!]
	\centering
	\begin{tabular}{cccc}
		\hline
		Strange quark mass & Coupling parameter & Maximum mass & Radius\\
		$(m_{s})~(MeV)$ & $(\alpha_{c})$ & $(M_{\odot})$ & $(Km)$ \\ \hline
		\multirow{5}{*}{100} & -2.0 & 2.059 & 10.99\\
		& -1.0 & 1.91 & 10.30 \\
		& 0 & 1.78 & 9.72 \\
		& 1.0 & 1.66 & 9.21 \\
		& -2.0 & 1.56 & 8.76 \\ \hline   
	\end{tabular}
	\caption{Maximum mass-radius for for $B=70~MeV/fm^{3}$ and $m_{s}=100~MeV$.\label{tab3}}
\end{table} 
\newpage 
\section{Physical features of the model under $f(R,T)$ gravity}\label{sec8} In this section, we have analysed the physical characteristic attributes of a compact object, namely, the energy density $(\rho)$, isotropic pressure $(p)$ for different suitable combinations of strange quark mass $(m_{s})$ and coupling parameter $(\alpha_{c})$ within the framework of $f(R,T)$ theory of gravity. For the analysis, we have considered EXO 1745-248 having mass $1.4~M_{\odot}$ \cite{Ozel}. We have predicted the radius of EXO 1745-248 using eq.~\eqref{eq35} for different representative values of $m_{s}$ and $\alpha_{c}$ and they are tabulated in tables~(\ref{tab4}) and (\ref{tab5}). It is noted from tables~(\ref{tab4}) and (\ref{tab5}) that the radius decreases when $m_{s}$ increases keeping $\alpha_{c}$ fixed and also the radius radius decreases when $\alpha_{c}$ increases keeping $m_{s}$ fixed.  
\begin{table}[ht!]
	\centering
	\begin{tabular}{ccc}
		\hline
		Coupling parameter & Strange quark mass & Predicted Radius (R)\\
		$(\alpha_{c})$ & $(m_{s})~(MeV)$ & $(Km)$ \\ \hline
		\multirow{4}{*}{-1.0} & 0 & 10.80 \\
		& 50 & 10.74 \\
		& 100 & 10.62 \\
		& 150 & 10.48 \\ 
		\hline 
	\end{tabular}
	\caption{Prediction of radius for EXO 1745-248 for $B=70~MeV/fm^{3}$ and $\alpha_{c}=-1.0$.\label{tab4}}
\end{table} 
\begin{table}[ht!]
	\centering
	\begin{tabular}{ccc}
		\hline
		Strange quark mass & Coupling parameter & Predicted Radius (R)\\
		$(m_{s})~(MeV)$ & $(\alpha_{c})$ & $(Km)$ \\ \hline
		\multirow{5}{*}{100} & -2.0 & 11.18 \\
		& -1.0 & 10.62 \\
		& 0 & 10.15 \\
		& 1.0 & 9.74 \\
		& 2.0 & 9.39 \\
		\hline 
	\end{tabular}
	\caption{Prediction of radius for EXO 1745-248 for $B=70~MeV/fm^{3}$ and $m_{s}=100$.\label{tab5}} 
\end{table}
\begin{table}[htbp]
	\centering
	\begin{tabular}{ccccccc}
		\hline
		\multirow{3}{*}{Compact objects} &&&&& \multicolumn{2}{c}{Predicted radius}\\ 
		& Measured mass & Measured radius & $\alpha_{c}$ & $m_{s}$ & \multicolumn{2}{c}{(Km)} \\ \cline{6-7}
		& $(M_{\odot})$ & (Km) & & $(MeV)$ & Model & TOV \\ 
		\hline
		\vspace{0.2cm}
		EXO 1745-248 \cite{Ozel} & 1.4 & 11 & -1.7 & 100 & 11.00 & 10.83 \\
		\vspace{0.2cm}
		4U 1820-30 \cite{Guver} & $1.58^{+0.06}_{-0.06}$ & $9.1^{+0.4}_{-0.4}$ & 1.5 & 150 & 9.75 & 8.91\\ 
		\vspace{0.2cm}
		LMC X-4 \cite{Rawls} & $1.04^{+0.09}_{-0.09}$ & $8.301^{+0.2}_{-0.2}$ &  2.5 & 150 & 8.38 & 8.35 \\ 
		\vspace{0.2cm}
		HER X-1 \cite{Abubekerov} & $0.85^{+0.15}_{-0.15}$ & $8.1^{+0.41}_{-0.41}$ & 1.5 & 150 & 8.18 & 8.20 \\
		\vspace{0.2cm}
		PSR J1614-2230 \cite{Demorest} & $1.97^{+0.04}_{-0.04}$ & 11-15 & -2.0 & 100 & 12.30 & 11.44 \\
		\vspace{0.2cm}
		4U 1608-52 \cite{Guver1} & $1.74^{+0.14}_{-0.14}$ & $9.3^{+1.0}_{-1.0}$ & 0.5 & 50 & 10.67 & 9.36 \\
		\vspace{0.2cm}
		PSR J0740+6620 \cite{Riley} & $2.072^{+0.067}_{-0.066}$ & $12.39^{-1.30}_{-0.98}$ & -2.5 & 100 & 12.85 & 11.81 \\
		\hline			
	\end{tabular}
	\caption{Tabulation of radius prediction of few known compact objects from (i) present model and (ii) TOV Mass-Radius relation.\label{tab6}} 
\end{table}  
\begin{table}[htbp]
	\centering
	\begin{tabular}{cccc}
		\hline
		Compact object & Central density $(\rho_{0})$ & Surface density $(\rho_{surface})$ & Central pressure $(p_{0})$ \\ 
		& $(g/cm^{3})$ & $(g/cm^{3})$ & $(dyn/cm^{2})$ \\ \hline
		EXO 1745-248 & $0.83\times10^{15}$ & $0.52\times10^{15}$ & $0.92\times10^{35}$\\
		4U 1820-30 & $1.01\times10^{15}$ & $0.54\times10^{15}$ & $1.43\times10^{35}$ \\
		LMC X-4 & $0.86\times10^{15}$ & $0.54\times10^{15}$ & $0.97\times10^{35}$ \\
		HER X-1 & $0.79\times10^{15}$ & $0.54\times10^{15}$ & $0.75\times10^{35}$ \\
		PSR J1614-2230 & $0.95\times10^{15}$ & $0.52\times10^{15}$ & $1.28\times10^{35}$\\
		4U 1608-52 & $0.94\times10^{15}$ & $0.50\times10^{15}$ & $1.33\times10^{35}$ \\
		PSR J0740+6620 & $0.95\times10^{15}$ & $0.52\times10^{15}$ & $1.28\times10^{35}$\\
		\hline
	\end{tabular}
	\caption{Tabulation of physical parameters using radius prediction from the model.\label{tab7}}
\end{table}
\begin{figure}[ht!]
	\centering
	\includegraphics[width=0.5\textwidth]{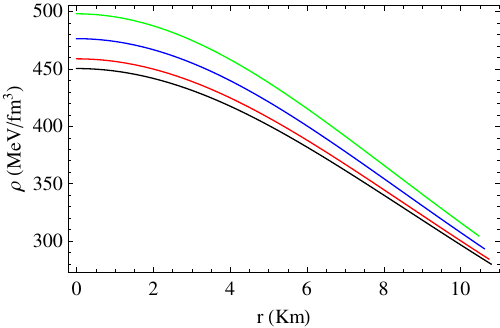}
	\caption{Radial variation of energy density $(\rho)$ for different $m_{s}$ keeping $B=70~MeV/fm^{3}$ and $\alpha_{c}=-1.0$. The black, red, blue and green lines correspond to $m_{s}=0,~50,~100,~150~MeV$ respectively.}
	\label{fig4} 
\end{figure}
\begin{figure}[ht!]
	\centering
	\includegraphics[width=0.5\textwidth]{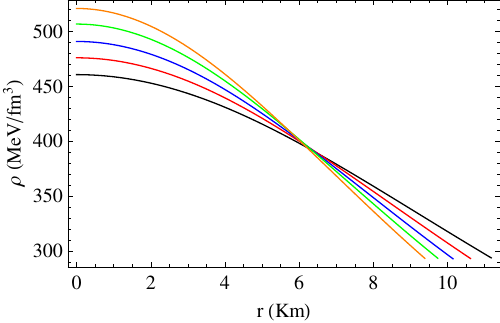}
	\caption{Radial variation of energy density $(\rho)$ for different coupling constant $\alpha_{c}$ keeping $B=70~MeV/fm^{3}$ and $m_{s}=100~MeV$. The black, red, blue, green and orange lines correspond to $\alpha_{c}=-2.0,~-1.0,~0,~1.0, ~2.0$ respectively.}
	\label{fig5}
\end{figure}
\begin{figure}[ht!]
	\centering
	\includegraphics[width=0.5\textwidth]{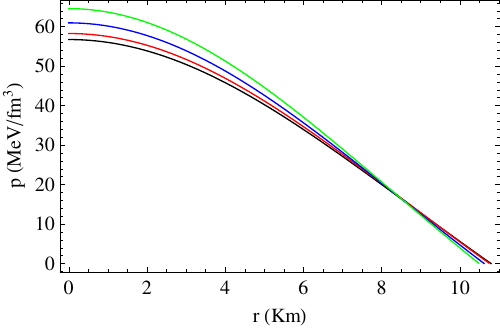}
	\caption{Radial variation of pressure $(p)$ for different $m_{s}$ keeping $B=70~MeV/fm^{3}$ and $\alpha_{c}=-1.0$. The black, red, blue and green lines correspond to $m_{s}=0,~50,~100,~150~MeV$ respectively.}
	\label{fig6} 
\end{figure}
\begin{figure}[ht!]
	\centering
	\includegraphics[width=0.5\textwidth]{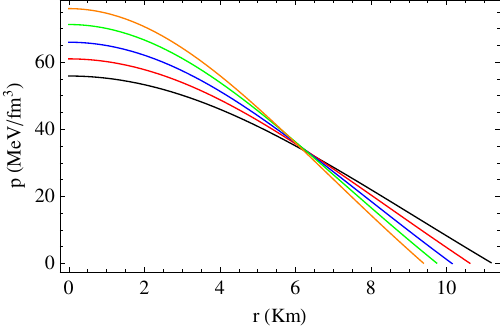}
	\caption{Radial variation of pressure $(p)$ for different coupling constant $\alpha_{c}$ keeping $B=70~MeV/fm^{3}$ and $m_{s}=100~MeV$. The black, red, blue, green and orange lines correspond to $\alpha_{c}=-2.0,~-1.0,~0,~1.0, ~2.0$ respectively.}
	\label{fig7} 
\end{figure}
From figures~(\ref{fig4}, \ref{fig5}, \ref{fig6}, \ref{fig7}), we note that, the energy density and pressure show a monotonically decreasing nature, which is a realistic feature for a stellar model in stable equilibrium. It must also be noted that, $(\frac{d\rho}{dr})<0$ and $(\frac{dp}{dr})<0$ in the interior of the stellar structure and $(\frac{d\rho}{dr})_{r=0}=0=(\frac{dp}{dr})_{r=0}$ as shown in figure~(\ref{fig8}). In addition to this, it is observed that at the stellar centre $(\frac{d^{2}\rho}{dr^{2}})<0$ and $(\frac{d^{2}p}{dr^{2}})<0$ which are shown in figure~(\ref{fig9}).
\begin{figure}[ht!]
	\hspace{-0.4cm}
	\begin{minipage}{.5\textwidth}
		\centering
		\includegraphics[width=1\linewidth]{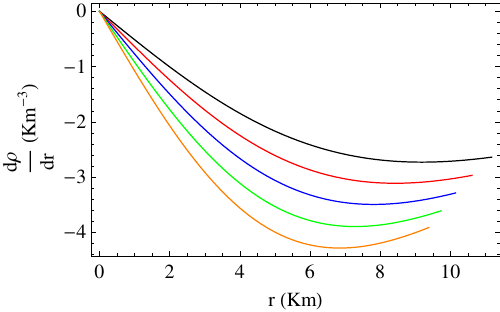}
	\end{minipage}%
	\begin{minipage}{.5\textwidth}
		\hspace{0.2cm}
		\includegraphics[width=1\linewidth]{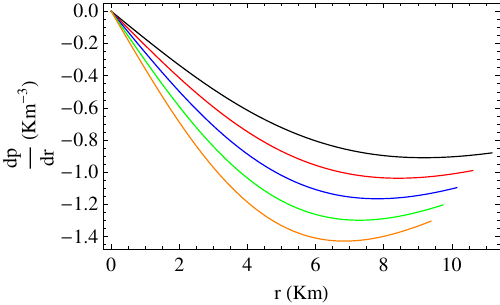}
	\end{minipage}
	\caption{Radial variation of $(\frac{d\rho}{dr})$ $(\times10^{-5})$ and $(\frac{dp}{dr})$ $(\times10^{-5})$ for different coupling constant $\alpha_{c}$ keeping $B=70~MeV/fm^{3}$. The  black, red, blue, green and orange lines correspond to $\alpha_{c}=-2.0,~-1.0,~0,~1.0, ~2.0$ respectively.}
	\label{fig8} 
\end{figure}
\begin{figure}[ht!]
	\hspace{-0.4cm}
	\begin{minipage}{.5\textwidth}
		\centering
		\includegraphics[width=1\linewidth]{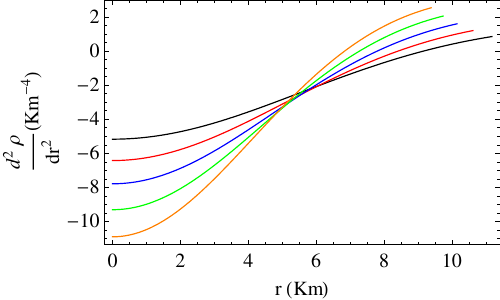}
	\end{minipage}%
	\begin{minipage}{.5\textwidth}
		\hspace{0.2cm}
		\includegraphics[width=1\linewidth]{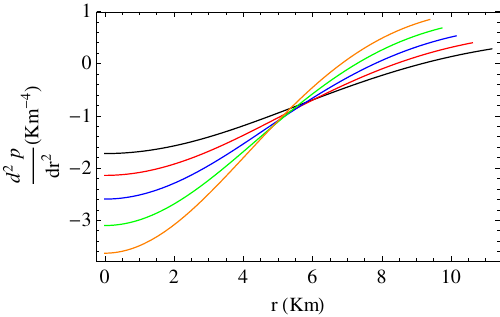}
	\end{minipage}
	\caption{Radial variation of $(\frac{d^{2}\rho}{dr^{2}})$ $(\times10^{-6})$ and $(\frac{d^{2}p}{dr^{2}})$ $(\times10^{-6})$for different coupling constant $\alpha_{c}$ keeping $B=70~MeV/fm^{3}$. The  black, red, blue, green and orange lines correspond to $\alpha_{c}=-2.0,~-1.0,~0,~1.0, ~2.0$ respectively.}
	\label{fig9}
\end{figure}
\newpage
\subsection{Causality Condition} To ensure a realistic modeling, the characterisation of internal dense matter is necessary. An efficient way of doing this is through the exploration of sound wave velocities. In an isotropic case, the expression for sound velocity reads, $v_{s}^{2}=\frac{dp}{d\rho}$, where $p$ and $\rho$ are the pressure and energy density respectively. By considering the system of units as $h=c=1$, the causality condition introduces an upper limit on the sound velocity as $v_{s}^{2}\leq1$. In the context of thermodynamic stability, $v_{s}^{2}>0$. Hence, within the stellar configuration, the cumulative condition of $0<v_{s}^{2}\leq1$ must hold simultaneously. Using eqs.~\eqref{eq27} and \eqref{eq28}, we obtain: 
\begin{equation}
	v_{s}^{2}=\frac{dp}{d\rho}=\frac{1}{3}. \label{eq40}
\end{equation}
Therefore, for the present isotropic model, the causality condition is satisfied throughout the interior of the stellar model irrespective of the presence of $m_{s}$ and $\alpha_{c}$. 
\subsection{Energy Condition} In the theory of gravity, the viability of a physically realistic energy-momentum tensor is ensured through the energy conditions. These conditions pave the way to understand the nature of the interior matter distribution without explicitly specifying the matter content. Therefore, it is possible to acquire physical attributes about extreme gravitational events such as gravitational collapse or geometrical singularity etc. without focusing on the energy density or pressure. In literature, the inspection of these energy conditions are an algebraic problem \cite{Kolassis} and more precisely the eigenvalue problem of the energy-momentum tensor. Exploring the energy conditions in a 4-dimensional space-time leads to 4-degree polynomial roots which is complicated as there are analytical solutions of eigenvalues. Despite the difficulty in obtaining a generalised solution, a viable and physically realistic matter distribution must follow the null (NEC), weak (WEC), strong (SEC) and dominant (DEC) energy conditions \cite{Kolassis,Hawking,Wald} within the stellar boundary. Here, in this section, we have studied the energy conditions \cite{Brassel, Brassel1} and found that the conditions are well satisfied in our present model. The present study includes the following inequalities that must hold simultaneously in presence of $m_{s}$, $B$ and $\alpha_{c}$: 
\begin{itemize}
	\item	NEC:~~~~$T_{ij}l^{i}l^{j}\geq0\Rightarrow\rho+p\geq0$, \label{eq41}\\
	\item	WEC:~~~~ $T_{ij}t^{i}t^{j}\geq0\Rightarrow\rho\geq0,~\rho+p\geq0$,  \label{eq42}\\
	\item	SEC:~~~~$T_{ij}t^{i}t^{j}-\frac{1}{2}T_{k}^{k}t^{\sigma}t_{\sigma}\geq0\Rightarrow\rho+\sum p\geq0,~or,~\rho+3p\geq0,$ \label{eq43} \\
	\item	DEC:~~~~$T_{ij}t^{i}t^{j}\geq0\Rightarrow\rho\geq|p|$, \label{eq44}
\end{itemize} 
where, $l^{i}$ and $t^{i}$ are time-like and null vectors respectively.  
\begin{figure}[ht!]
	\centering
	\includegraphics[width=0.5\textwidth]{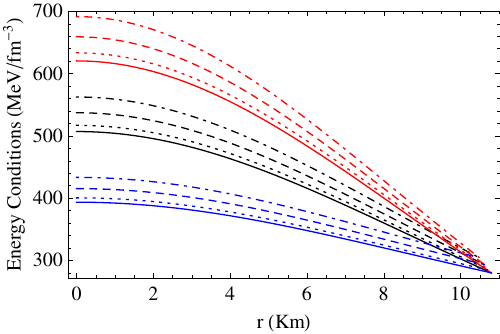}
	\caption{Radial variation of energy conditions for different $m_{s}$ keeping $B=70~MeV/fm^{3}$ and $\alpha_{c}=-1.0$. The black, red and blue lines represent $(\rho+p)$, $(\rho+3p)$ and $(\rho-p)$ respectively and the solid, dotted, dashed and dot-dashed lines correspond to $m_{s}=0,~50,~100,~150~MeV$ respectively.}
	\label{fig10} 
\end{figure}
\begin{figure}[ht!]
	\centering
	\includegraphics[width=0.5\textwidth]{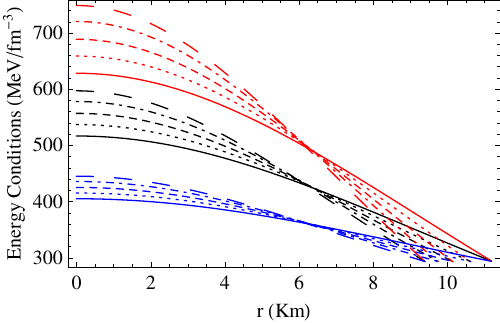}
	\caption{Radial variation of energy conditions for different coupling parameter $\alpha_{c}$ keeping $B=70~MeV/fm^{3}$ and $m_{s}=100~MeV$. The black, red and blue lines lines represent $(\rho+p)$, $(\rho+3p)$ and $(\rho-p)$ respectively and the solid, dotted, dashed, dot-dashed and large dashed lines correspond to $\alpha_{c}=-2.0,~-1.0,~0,~1.0, ~2.0$ respectively.}
	\label{fig11} 
\end{figure}
\newpage
\section{Stability Analysis} \label{sec9} We have studied the stability of the present model through the following criteria: \\
(i) Generalised TOV equation, \\
(ii) Radial variation of adiabatic index, \\
(iii) Lagrangian perturbation and \\
(iv) Tidal love number and Tidal deformability 
\subsection{Generalised TOV equation} 
\begin{figure}[ht!]
	\centering
	\includegraphics[width=0.5\textwidth]{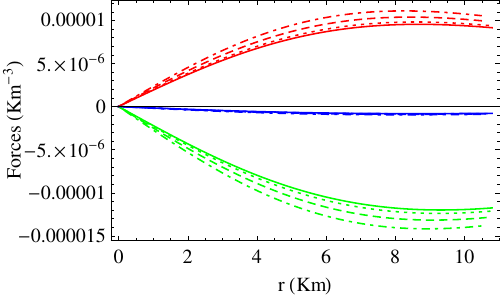}
	\caption{Radial variation of different forces for different $m_{s}$ keeping $B=70~MeV/fm^{3}$ and $\alpha_{c}=-1.0$. The red, blue and green lines represent $F_{h}$, $F_{m}$ and $F_{g}$ respectively and the solid, dotted, dashed and dot-dashed lines correspond to $m_{s}=0,~50,~100,~150~MeV$ respectively.}
	\label{fig12} 
\end{figure}

\begin{figure}[ht!]
	\centering
	\includegraphics[width=0.5\textwidth]{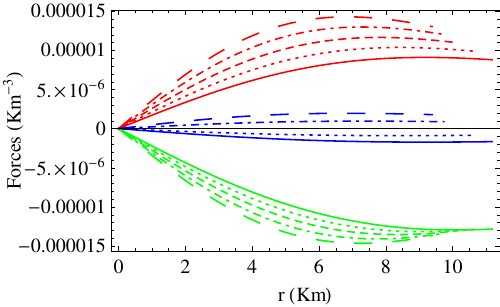}
	\caption{Radial variation of different forces for different $m_{s}$ keeping $B=70~MeV/fm^{3}$ and $m_{s}=100~MeV$. The red, blue and green lines represent $F_{h}$, $F_{m}$ and $F_{g}$ respectively and the solid, dotted, dashed, dot-dashed and large dashed lines correspond to $\alpha_{c}=-2.0,~-1.0,~0,~1.0, ~2.0$ respectively.}
	\label{fig13} 
\end{figure}

The stability of compact stellar objects immensely rely on the equilibrium configuration of a model under the influence of different forces throughout the object. In the present scenario, we are dealing with an isotropic model in the modified $f(R,T)$ gravity and the study of stability is based on the following force components- (i) the gravitational force $(F_{g})$, (ii) the hydrostatic force $(F_{h})$ and (iii) the additional contribution due to the modified gravity effects $(F_{m})$. Formulation of the generalised Tolman-Oppenheimer-Volkoff (TOV) equation \cite{Tolman,Oppenheimer} in this context is expressed as : 
\begin{equation}
	-\frac{M_{G}(r)(\rho+p)}{r^2}e^{\lambda-\nu}-\frac{dp}{dr}+\frac{\alpha_{c}}{8\pi+2\alpha_{c}}(p'-\rho')=0, \label{eq45}
\end{equation} 
where, the active gravitational mass $(M_{G})$ derived from the Tolman-Whittaker \cite{Gron} mass formula is given as: 
\begin{equation}
	M_{G}(r)=r^{2}\nu'e^{\nu-\lambda}. \label{eq46}
\end{equation}
Substituting eq.~\eqref{eq46} in eq.~\eqref{eq45}, we obtain
\begin{equation}
	F_{g}+F_{h}+F_{m}=0. \label{eq47}
\end{equation}  
Here, 
\begin{equation}
	F_{g}=-\nu'(\rho+p),  \label{eq48} 
\end{equation}
\begin{equation}
	F_{h}=-\frac{dp}{dr},  \label{eq49}
\end{equation}	
and
\begin{equation}
	F_{m}=\frac{\alpha_{c}}{8\pi+2\alpha_{c}}(p'-\rho').  \label{eq50}
\end{equation}
Using eqs.~\eqref{eq27}, \eqref{eq28}, \eqref{eq29}, we can compute the necessary expressions for the set of equations eqs.~\eqref{eq48}, \eqref{eq49}, \eqref{eq50}. To avoid the complexity of mathematical expressions, we have shown the equilibrium conditions through a graphical representations in figures~(\ref{fig12}) and (\ref{fig13}). 
\newpage
\subsection{Radial variation of adiabatic index} The adiabatic index for a relativistic isotropic stellar configuration is expressed as: 
\begin{equation}
	\Gamma=\frac{\rho+p}{p}\frac{dp}{d\rho}. \label{eq51}
\end{equation}
Heintzmann and Hillebrandt \cite{Heintzmann} in their study showed that for a stable general relativistic isotropic fluid sphere, the adiabatic index must be greater than its Newtonian limit of $\frac{4}{3}$, i.e., $\Gamma>\frac{4}{3}$. 
\begin{figure}[ht!]
	\centering
	\includegraphics[width=0.5\textwidth]{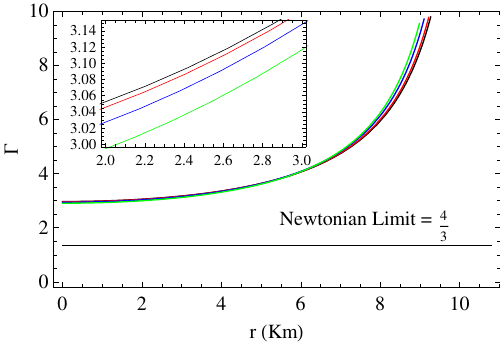}
	\caption{Radial variation of adiabatic index $(\Gamma)$ for different $m_{s}$ keeping $B=70~MeV/fm^{3}$ and $\alpha_{c}=-1.0$. The black, red, blue and green lines correspond to $m_{s}=0,~50,~100,~150~MeV$ respectively.}
	\label{fig14} 
\end{figure}
\begin{figure}[ht!]
	\centering
	\includegraphics[width=0.5\textwidth]{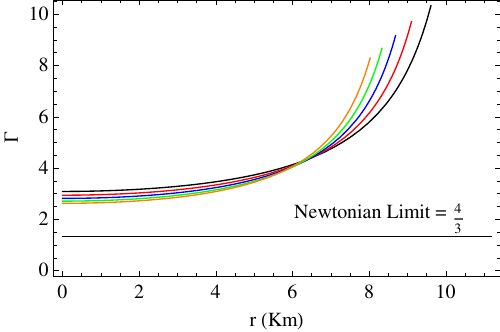}
	\caption{Radial variation of adiabatic index $(\Gamma)$ for different $m_{s}$ keeping $B=70~MeV/fm^{3}$ and $m_{s}=100~MeV$. The black, red, blue, green and orange lines correspond to $\alpha_{c}=-2.0,~-1.0,~0,~1.0, ~2.0$ respectively.}
	\label{fig15} 
\end{figure}
\\Hence, from figures~(\ref{fig14}) and (\ref{fig15}), we note that the variations of adiabatic index represent a stable stellar model.
\newpage
\subsection{Lagrangian perturbation} We study the stability of our present model under the impact of small radial oscillations through the Lagrangian change of pressure at the surface of the compact object with frequency $(\omega^{2})$. According to Pretel \cite{Pretel1}, the Lagrangian perturbation depends on the frequency and following that procedure, we express the coupled equations for radial oscillations as:
\begin{equation}
	\frac{d\zeta}{dr}=-\frac{1}{r}(3\zeta+\frac{\Delta p}{\Gamma p})+\frac{d\nu}{dr}\zeta,\label{eq52} 
\end{equation}
\begin{equation} 
	\frac{d\Delta p}{dr}=\zeta\Big(\frac{\omega^2}{c^2}e^{2(\lambda-\nu)}(\rho+p)r-4\frac{dp}{dr}-\frac{8\pi G}{c^4}(\rho+p)e^{2\lambda}rp+r(\rho+p)(\frac{d\nu}{dr})\Big)-\Delta p\Big(\frac{d\nu}{dr}+\frac{4\pi G}{c^4}(\rho+p)re^{2\lambda}\Big), \label{eq53}
\end{equation} 
where, $\zeta~(=\frac{\delta(r)}{r})$ represents the radial eigen function of the Lagrangian displacement. In this formalism, $\zeta$ is normalised such that $\zeta(0)=1$. The term with $\Big(\frac{1}{r}\Big)$ should vanish to avoid the central singularity, when $r\rightarrow0$, appearing in eq.~\eqref{eq52}. Hence, we arrive at the condition: 
\begin{equation}
	\Delta p=-3\Gamma\zeta p. \label{eq54}
\end{equation} 
At the stellar boundary $(r=R)$, the Lagrangian change in pressure must also vanish, i.e.,
\begin{equation}
	\Delta p=0. \label{eq55}
\end{equation}
\begin{figure}[h!]
	\centering
	\includegraphics[width=0.5\textwidth]{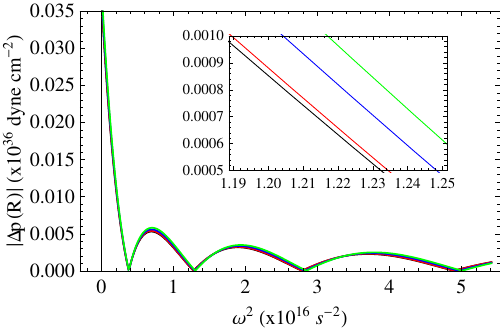}
	\caption{Radial variation of adiabatic index $(\Gamma)$ for different $m_{s}$ keeping $B=70~MeV/fm^{3}$ and $\alpha_{c}=-1.0$. The black, red, blue and green lines correspond to $m_{s}=0,~50,~100,~150~MeV$ respectively.}
	\label{fig16} 
\end{figure}
\begin{figure}[ht!]
	\centering
	\includegraphics[width=0.5\textwidth]{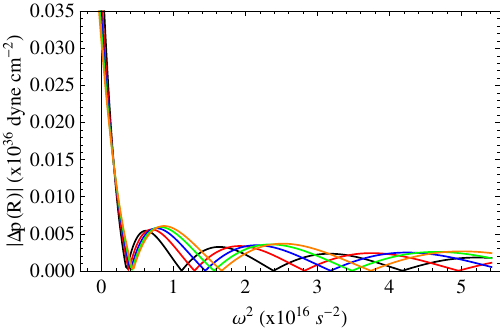}
	\caption{Radial variation of adiabatic index $(\Gamma)$ for different $m_{s}$ keeping $B=70~MeV/fm^{3}$ and $m_{s}=100~MeV$. The black, red, blue, green and orange lines correspond to $\alpha_{c}=-2.0,~-1.0,~0,~1.0, ~2.0$ respectively.}
	\label{fig17} 
\end{figure}
From figures~(\ref{fig16}) and (\ref{fig17}), we note that $\omega^{2}>0$ for all the normal modes of radial oscillations in the present model. Also, the minima of the plots represent the correct values of the normal frequency. Hence, we may assert that our model is also stable under the influence of small radial perturbations within the parameter space used here.
\newpage
\subsection{Tidal Love Number and Tidal deformability} Tidal Love Number (TLN) is a parameter that measures the distortion of the spherical shape of a compact stellar body due to the external perturbing effects of gravity. It is generally denoted by $k_{2}$. Similarly, Tidal deformability $(\Lambda)$ defines the surface disruption. $k_{2}$ and $\Lambda$ are related by the following expression: 
\begin{equation}
	k_{2}=\frac{3}{2}\Lambda\Big(\frac{M}{R}\Big)^{5}, \label{eq56}
\end{equation} 
where $M$ is the mass and $R$ is the radius of the compact object. Following Refs. \cite{Hinderer,Hinderer1}, the expression for TLN with even parity $(l=2)$ is written as:
\begin{eqnarray}
	\hspace{-1cm}	k_{2}=\frac{8u^{5}}{5}(1-2u)^{2}[2+2u(g-1)-g]\times\Big[2u[6-3g+3u(5g-8)]\nonumber\\+4u^{3}[13-11g+u(3g-2)+2u^{2}(1+g)]+3(1-2u)^{2}[2-g+2u(g-1)log(1-2u)]\Big]^{-1}, \label{eq57}
\end{eqnarray} 
where, $u=\frac{M}{R}$ is the compactness and $g=\frac{RH'(R)}{H(R)}$ and $H(R)=H(r)|_{r\rightarrow R}$. The prime denotes the derivative with respect to $r$. Now, $H(r)$ is expressed as: 
\begin{equation}
	H(r)=c_{1}\Big(\frac{r}{M}\Big)^{2}\Big(1-\frac{2M}{r}\Big)\Big[-\frac{M(M-r)(2M^{2}+6Mr-3r^{2})}{r^{2}(2M-r)^{2}}+\frac{3}{2}log\Big(\frac{r}{r-2M}\Big)\Big]+3c_{2}\Big(\frac{r}{M}\Big)^{2}\Big(1-\frac{2M}{r}\Big). 
\end{equation} 
Abbott and Abbott \cite{Abbott} have predicted some constraints imposed on $\Lambda$ by the event GW170817. Following the article \cite{Bauswein}, the Tidal deformability for a star having a mass of $1.4~M_{\odot}$ should obey the limit, $\Lambda<800$. 
\begin{table}[h]
	\centering 
	\begin{tabular}{ccccc}
		\hline
		Coupling & Strange quark mass & Predicted & $k_{2}$ & $\Lambda$ \\
		parameter & $(m_{s})~(MeV)$ & radius (R) & & \\
		$(\alpha_{c})$ &  & $(Km)$ & & \\ \hline
		\multirow{4}{*}{-1.0} & 0 & 10.80 & 0.02855566 & 74.50 \\ 
		& 50 & 10.74 & 0.0285525 & 72.44 \\ 
		& 100 & 10.62 & 0.0284549 & 68.25 \\
		& 150 & 10.48 & 0.0283703 & 63.67 \\ \hline
	\end{tabular}
	\caption{Tidal deformability of our present model for $B=70~MeV/fm^{3}$ and $\alpha_{c}=-1.0$.\label{tab8}}
\end{table} 
\begin{table}[h]
	\centering 
	\begin{tabular}{ccccc}
		\hline
		Strange quark mass & Coupling & Predicted & $k_{2}$ & $\Lambda$ \\
		$(m_{s})~(MeV)$ & parameter & radius (R) & & \\
		& $(\alpha_{c})$ & $(Km)$ & & \\ \hline
		\multirow{5}{*}{100} & -2.0 & 11.18 & 0.0317997 & 98.61  \\ 
		& -1.0 & 10.62 & 0.0284549 & 68.25 \\ 
		& 0 & 10.15 & 0.0260266 & 49.78 \\ 
		& 1.0 & 9.74 & 0.0241767 & 37.63 \\ 
		& 2.0 & 9.39 & 0.0226739 & 29.39 \\ \hline
	\end{tabular}
	\caption{Tidal deformability of our present model for $B=70~MeV/fm^{3}$ and $m_{s}=100~MeV$.\label{tab9}}
\end{table}
\\Tables~(\ref{tab2}) and (\ref{tab3}) shows that as the strange quark mass $(m_{s})$ increases, the maximum mass of a compact stellar object decreases. Similarly, as the gravity-matter coupling becomes stronger, the maximum mass decreases. From eq.~\eqref{eq56}, we note that Tidal deformability depends on the compactness $(u=\frac{M}{R})$. In the present model, the radius prediction shows that as we increase $m_{s}$ and $\alpha_{c}$, the radius of decreases when the mass is kept constant at $M=1.4~M_{\odot}$, which results in the increase of compactness. Therefore, with the increasing compactness, the compressibility of the stellar configuration due to gravity decreases. The tidal deformability decreases in the present model.    
\section{Discussion} \label{sec10} In this paper, we have studied the effect of strange quark mass $(m_{s})$ and coupling constant $(\alpha_{c})$ on the formulation of isotropic strange compact stars within the framework of modified $f(R,T)$ gravity theory. To obtain a compatible stellar model, we have used the Buchdahl-I metric ansatz \cite{Buchdahl} along with a specific class of $f(R,T)$ models {\it viz.} $f(R,T)=R+2\alpha_{c}T$ \cite{Harko1}, where $\alpha_{c}$ is termed as the gravity-matter coupling constant. Considering the above specifications and using MIT bag model EoS in presence of strange quark mass $(m_{s})$ \cite{Kapusta}, we have set up the EFE and obtained an exact set of solutions for the present model. Considering the quark star hypothesis, we have calculated the energy per baryon number to constrain the strange quark mass $(m_{s})$ within the stable region, i.e., $\mathcal{E}_{B}<930.4~MeV$ $(^{56}Fe)$ and we have noted that a bag constant $B=70~MeV/fm^{3}$ ensures a stable modeling when $m_{s}$ is varied upto $150~MeV$. To accommodate compact stars from low mass to high mass region, we have considered the range of $\alpha_{c}$ from -2.0 to 2.0 following eq.~\eqref{eq39}. Using the range of $\alpha_{c}$ and variation of $m_{s}$, we have numerically solved the TOV equations \cite{Tolman,Oppenheimer} to obtain the maximum mass and the corresponding radius in the model. It is noted that $\alpha_{c}$ and $m_{s}$ have some distinctive effects on the maximum mass of the model and they are tabulated in tables~(\ref{tab2}) and (\ref{tab3}). From the calculation of maximum mass and radius prediction in the present model, we have noted that, the model is suitable in explaining the observed radii of low mass compact SS candidates for positive values of coupling constant $(\alpha_{c})$. However, if we consider the negative value of coupling constant $(\alpha_{c})$, we can determine and incorporate the compact stars having high mass and radii. In the cosmological scenario, negative values of $\alpha_{c}$ signify a transition from deceleration to accelerating phase of the universe \cite{Velten}. In the astrophysical context, we may hypothesise that this negative $\alpha_{c}$ defines a repulsive coupling which reduces the compressibility of the compact stars and enters a weak gravity regime. This in turn results in a stiffer EoS, i.e., the maximum mass and radius increase. Considering the lower limit of eq.~\eqref{eq39}, we have checked that the numerical solution of TOV equation \cite{Tolman,Oppenheimer} for $\alpha_{c}=-3.7$ and $m_{s}=100~MeV$ yield a maximum mass of $2.35~M_{\odot}$ which corresponds to the fastest and heaviest pulsar PSR J0952-0607 \cite{Carvalho2} in the Milky Way disk. Similarly, $\alpha_{c}=-4.9$ and $m_{s}=100~MeV$ yield a maximum mass of $2.59~M_{\odot}$ which describes the lighter component of GW 190814 \cite{Abbott1}. Therefore, to explain the compact stars in high mass region, the negative values of $\alpha_{c}$ may be considered. Using the boundary conditions, we have found an expression for predicting the radii for different compact objects as given in  eq.~\eqref{eq35}. To ensure a physically realistic modeling, we have considered EXO 1745-248 having mass $1.4~M_{\odot}$ \cite{Ozel} and predicted its radius following eq.~\eqref{eq35} with systematic variations of $m_{s}$ and $\alpha_{c}$ and they are tabulated in tables~(\ref{tab4}) and (\ref{tab5}). Using the radii, we have analysed the radial variations of the compact star properties namely, energy density, pressure and their gradients in figures~(\ref{fig4})-(\ref{fig9}) and found that they are well satisfied within the stellar interior. In table~(\ref{tab6}), we have shown the radii prediction for a wide range of compact SS through the prediction obtained from the model via eq.~\eqref{eq35} and TOV approach. We have noted that, the comparative results from both of the procedures are compatible with observed radii. A study of the sound velocity $v_{s}^{2}=(\frac{dp}{dr})$ reveals that $v_{s}^{2}=\frac{1}{3}$ in the model which satisfies the causality condition, i.e., $0\leq v_{s}^{2}\leq1$ and acts in favour of a stable stellar structure of strange quark stars according to the work of Bedaque and Steiner \cite{Bedaque} . We have noted that the energy conditions as shown in figures~(\ref{fig10}) and (\ref{fig11}) are well satisfied in the model. To analyse the stability of the model under the influence of different forces, we have studied their radial variations as shown in figures~(\ref{fig12}) and (\ref{fig13}). In figures~(\ref{fig14}) and (\ref{fig15}), we have shown the variation of adiabatic index and found that it is well above the Newtonian limit of $\frac{4}{3}$ which is a necessary criteria for an isotropic stable stellar configuration. To explore the stability of the present model under small radial perturbations we have plotted the absolute value of Lagrangian change in radial pressure at the stellar boundary with the eigen frequencies of different oscillation modes and we found real frequency spectrum $(\omega^{2}>0)$. To show the external effects of gravity on the present model, we have calculated the TLN and Tidal deformability for systematic variations of $\alpha_{c}$ and $m_{s}$. It is to be noted that $\Lambda<800$ for the present model which represents a stable configuration. Thus, we can infer that our current model effectively explains the mass-radius relationship and internal properties of a wide range of compact SQS candidates. Furthermore, it predicts a stable configuration for compact stars within the context of $f(R,T)$ theory of gravity.

\section{acknowledgments}
DB is thankful to DST for providing the fellowship vide no: DST/INSPIRE/03/2022/000001. We are also thankful to Jo\~ao Luis Rosa, Assistant Professor, University of Gdansk, Poland for the fruitful discussions on our manuscript.   



\end{document}